\documentclass[aps,prl,twocolumn,a4paper,showpacs]{revtex4}

\usepackage{amsfonts,amsmath,amssymb}
\usepackage{bm}\let\vec\bm
\usepackage{graphicx}
\usepackage{color}
\usepackage{hyperref}

\begin{document}

\newcommand{\Dp}{\Delta_p}
\newcommand{\Dpr}{\Delta_p(\vec{r})}
\newcommand{\Dpb}{\bar{\Delta}_p(\varphi)}
\newcommand{\Ed}{E_{\text{dip}}}
\newcommand{\Wd}{\Omega_{\text{dip}}}
\newcommand{\Wdr}{\Omega_{\text{dip}}(\vec{r})}
\newcommand{\Wdb}{\bar{\Omega}_{\text{dip}}(\varphi)}

\title{\boldmath Imaging the essential role of spin-fluctuations in high-$T_c$
superconductivity}

\author{N. Jenkins,$^{1}$ Y. Fasano,$^{1,2}$ C. Berthod,$^{1}$ I.
Maggio-Aprile,$^{1}$ A. Piriou,$^{1}$ E. Giannini,$^{1}$ B. W. Hoogenboom,$^{3}$
C. Hess,$^{4}$ T. Cren,$^{5}$ and \O. Fischer$^{1}$}

\affiliation{$^{1}$ DPMC-MaNEP, University of Geneva, 24 Quai Ernest-Ansermet,
1211 Geneva 4, Switzerland.}

\affiliation{$^{2}$ Instituto Balseiro and Centro At\'omico Bariloche, San Carlos de Bariloche, Argentina}

\affiliation{$^{3}$ London Centre for Nanotechnology and Department of Physics
and Astronomy, University College London, 17-19 Gordon Street, London WC1H 0AH,
United Kingdom.}

\affiliation{$^{4}$ Leibniz-Institute for Solid State and Materials Research,
IFW-Dresden, Dresden 01171, Germany.}

\affiliation{$^{5}$ Institut des Nano-Sciences de Paris, Universit\'{e}s Paris 6
et Paris 7 et C.N.R.S. (UMR 75 88), 140 rue de Lourmel, Campus Boucicaut, Paris
75015, France.}

\date{\today}

\begin{abstract}

We have used scanning tunneling spectroscopy to investigate short-length
electronic correlations in three-layer Bi$_{2}$Sr$_{2}$Ca$_{2}$Cu$_{3}$O$_{10 +
\delta}$ (Bi-2223). We show that the superconducting gap and the energy $\Wd$,
defined as the difference between the dip minimum and the gap, are both
modulated in space following the lattice superstructure, and are locally
anti-correlated. Based on fits of our data to a microscopic strong-coupling
model we show that $\Wd$ is an accurate measure of the collective mode energy in
Bi-2223. We conclude that the collective mode responsible for the dip is a local
excitation with a doping dependent energy, and is most likely the $(\pi,\pi)$
spin resonance.

\end{abstract}


\maketitle

The presence of phonon signatures in the electron tunneling spectra of classical
superconductors \cite{Giaever-1962}, and their quantitative explanation by the
Eliashberg equations \cite{Eliashberg-1960, McMillan-1965}, stand among the most
convincing validations of the BCS phonon-mediated pairing theory
\cite{Bardeen-1957}. For high-$T_{c}$ superconductors, the pairing mechanism
still remains an intriguing mystery. Several cuprate superconductors present a
spectroscopic feature, known as the dip-hump \cite{Huang-1989, Dessau-1991,
Renner-1995}, that resembles the phonon signatures of classical superconductors.
There is still no consensus on the origin of the dip-hump nor its connection to
high-$T_c$ superconductivity. In this work we report on a scanning tunneling
microscopy (STM) study of the three-layer compound
Bi$_{2}$Sr$_{2}$Ca$_{2}$Cu$_{3}$O$_{10 + \delta}$ (Bi-2223). We observe that the
gap magnitude, a direct measure of the pairing strength, is periodically
modulated on a lengthscale of about 5 crystal unit-cells. This variation follows
the superstructure, a periodic modulation of atomic positions naturally present
in Bi-based cuprates. By fitting the STM data with a strong-coupling model
\cite{Eschrig-2000, Hoogenboom-2003b} we demonstrate that the dip feature
originates from a collective excitation. This allows us to image the collective
mode energy (CME) at the atomic scale and reveal a modulation that also follows
the superstructure. The CME and the gap are locally anti-correlated. These
findings support that the collective mode probed in our study is related to
superconductivity, and is most likely the anti-ferromagnetic spin resonance
detected by neutron scattering \cite{Fong-1999}. Our results, in particular the
CME value of 30--40~meV, are in agreement with the spin-fluctuation-mediated
pairing scenario \cite{Monthoux-1994, Chubukov-2003}, in which the spin
resonance in high-$T_c$'s is a consequence of pairing.

An important issue in understanding high-$T_c$ superconductivity (HTS) is the
relevance of the observed inhomogeneous electronic properties to
superconductivity \cite{Fischer-2007}. The relationship between these
inhomogeneities and structural \cite{Slezak-2008} or dopant atom-
\cite{McElroy-2005, Chatterjee-2008} disorder is actively debated. Bi-based
cuprates are suitable materials to study this problem since they naturally
present a bulk modulation of atomic positions \cite{Gao-1988, Giannini-2008}.
Scanning tunneling microscopy (STM) images this superstructure as a
one-dimensional surface corrugation \cite{Fischer-2007}. A recent study on the
two-layer compound Bi$_{2}$Sr$_{2}$CaCu$_{2}$O$_{8+\delta}$ (Bi-2212) reports a
spatial modulation of the gap following the superstructure \cite{Slezak-2008}.
This phenomenon calls for an investigation on the sensitivity of other
spectroscopic features to variations of the interatomic distances. In
particular, since the dip feature can be regarded as the fingerprint of
Bogoliubov quasiparticles coupling with a collective excitation
\cite{Campuzano-1999, Eschrig-2000, Hoogenboom-2003b}, its spatial variation can
bear information on the nature of this interaction.

The existence of a dip in tunneling conductance at energies larger than the gap
was reported for several HTS \cite{Fischer-2007}. The dip is detected neither
above $T_c$ nor within the vortex cores \cite{Fischer-2007}, suggesting an
intimate connection with the superconducting state. We explore this possibility
by mapping the local tunneling conductance of
Bi$_{2}$Sr$_{2}$Ca$_{2}$Cu$_{3}$O$_{10 + \delta}$ (Bi-2223), the Bi-based
cuprate with the strongest dip \cite{Fischer-2007, Kugler-2006}. We find that
the local gap $\Dpr$ presents a one-dimensional modulation which follows the
superstructure, being largest at the maxima of the surface corrugation. We
observe that the energy separation between the dip and the coherence peak,
$\Wd$, is also modulated with the same period. Excellent fits of the STM spectra
with a model that considers a strong-coupling of quasiparticles with a
collective mode \cite{Eschrig-2000} establish that $\Wd$ is an accurate measure
of the CME. This allows us to reveal that the CME is locally anti-correlated
with the gap. As we will discuss, these findings support that the collective
mode is the $(\pi,\pi)$ spin resonance \cite{Fong-1999}.

\begin{figure}[tb]
\includegraphics[width=\columnwidth]{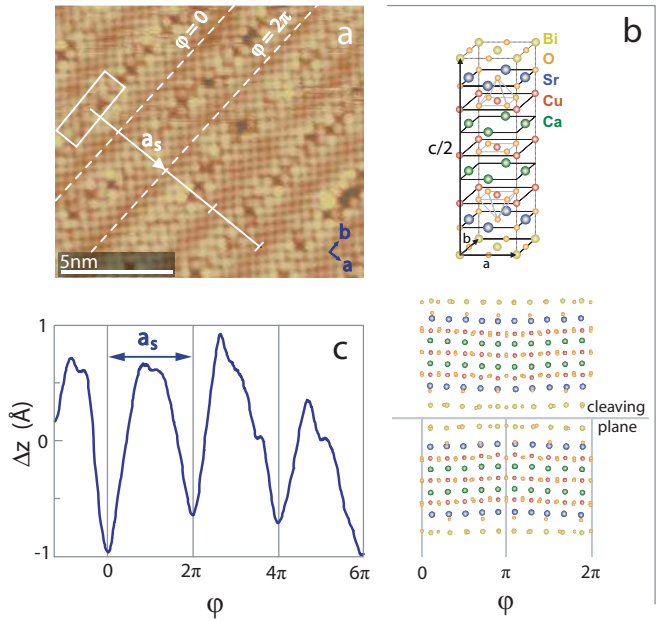}
\caption{\label{fig:fig1}
STM topography and crystal structure of Bi-2223. (a) Surface topography of an
underdoped sample ($T_c=103 \pm 2$~K) acquired at 5~K in constant-current mode
(tunneling current set to 0.2~nA and sample bias to 0.4~V). The Bi atoms exposed
after cleaving the sample are observed as bright spots. The in-plane unit cell
vectors of the ideal crystal structure, $\vec{a}$ and $\vec{b}$, and of the
superstructure, $\vec{a}_{s}$, are indicated. Lines of constant phase are
depicted [see also (b)]. The so-called missing atom row \cite{Kirk-1988} is
observed at $\varphi=(2n+1)\pi$. (b) Top panel: upper-half of the ideal Bi-2223
unit cell. The CuO bond direction is rotated 45$^{\circ}$ from the $\vec{a}$ and
$\vec{b}$ axes. Bottom panel: schematic representation of the periodic
modulation of atomic positions along the $\vec{c}$ axis. The in-plane location
is measured as a phase with $\varphi = 2n \pi$ corresponding to minima in the
vertical position of the Bi atoms of the layer exposed by cleaving. (c)
One-dimensional corrugation of the surface as a function of $\varphi$. The
profile was obtained by computing the average height within the box depicted in
(a), and by displacing this box along the white line.
}
\end{figure} 

The Bi-2223 compound investigated here presents the highest $T_c$ among the
Bi-based cuprates. Our crystals, free of Bi-2212 intergrowths, were grown by the
travelling floating zone method and oxygen-annealed \cite{Giannini-2008,
Piriou-2008}. We studied underdoped ($T_c=103 \pm 2$~K), nearly-optimally-doped
($T_c=109.5 \pm 0.5$~K) and overdoped ($T_c=108.5 \pm 0.5$~K) crystals using a
home-built UHV scanning-tunneling-microscope. The crystals were cleaved at room
temperature in a 1--$5\times10^{-9}$~mbar atmosphere and immediately cooled down
to 5~K. Iridium tips served as ground electrode and the bias voltage was applied
to the sample. The local tunneling conductance $dI/dV$ was obtained by numerical
derivation of the measured $I(V)$.

STM measurements in Bi-2223 image the Bi atoms of the topmost BiO plane exposed
by cleaving and the incommensurate modulation of the atomic positions (see
Fig.~\ref{fig:fig1}). The unit cell vector of this superstructure $\vec{a}_s$ is
parallel to the $\vec{a}$ axis---at 45$^{\circ}$ from the Cu-O bond direction.
For the nearly optimally-doped sample we estimate $a_{s}=27\pm 2$~\AA~$\approx 5
a$, in agreement with x-ray diffraction data \cite{Giannini-2008}. The maxima
and minima of the surface corrugation are nearly-straight lines perpendicular to
$\vec{a}_{s}$. We choose one of the lines of minima as the origin, and assign to
each point $\vec{r}$ a phase $\varphi$ proportional to the orthogonal distance
to this reference.

\begin{figure}[bt]
\centerline{\includegraphics[width=0.8\columnwidth]{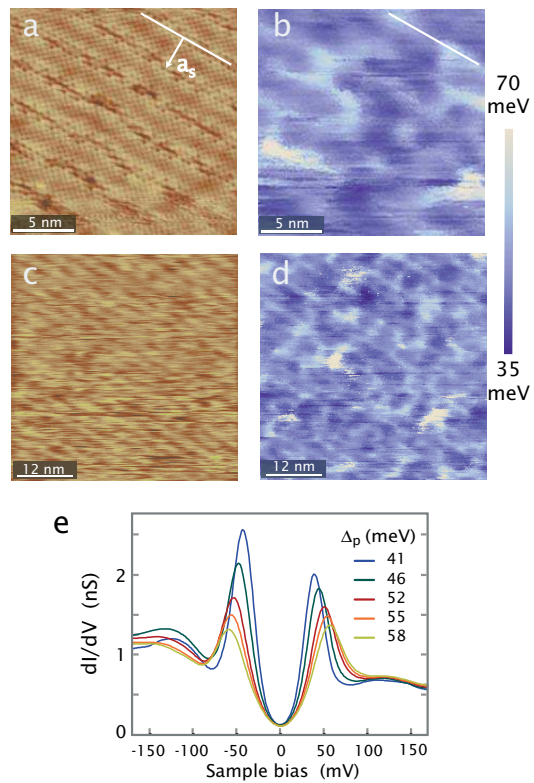}}
\caption{\label{fig:fig2}
Local gap modulation and average spectra for nearly optimally-doped Bi-2223. (a)
Topographic image of a $17\times17$~nm$^2$ region and (b) simultaneously
measured local gap map $\Dpr$. The white line is a spatial reference. (c)
Topographic image and (d) simultaneous local gap map in a $46\times46$~nm$^2$
field-of-view in a different region of the sample. The color scales in (b) and
(d) are identical and bright areas depict larger gap values. (e) Average
tunneling conductance spectra obtained from thousands of spectra with the same
gap acquired over the field-of-view of (b). Measurements were performed at 5~K.
}
\end{figure}

\begin{figure}[tb]
\centerline{\includegraphics[width=0.8\columnwidth]{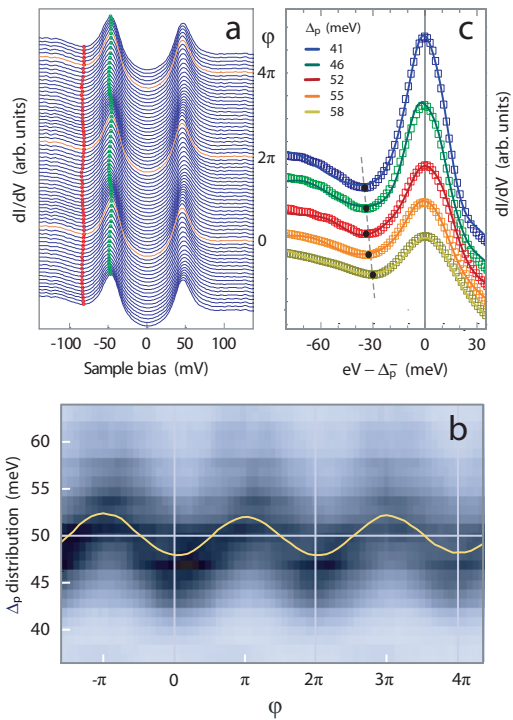}}
\caption{\label{fig:fig3}
Spatial modulation of the gap and dip energy in nearly optimally-doped Bi-2223.
(a) Evolution of the $dI/dV$ spectra with the phase $\varphi$: every curve
results from averaging thousands of spectra with the same $\varphi$. The spectra
measured at the minima of the surface corrugation are highlighted in orange. The
green and red dots indicate the coherence peaks and the dip feature for negative
bias, respectively. (b) Color-scale representation of $\Dp$ distributions for
spectra with the same $\varphi$: darker pixels correspond to more frequent gap
values. The yellow curve depicts the average gap as a function of $\varphi$,
$\bar{\Delta}_{p}(\varphi)$, computed from the geometrical mean of each
histogram. (c) Negative-bias part of the $\Dp$-averaged spectra shown in
Fig.~\ref{fig:fig2}c. Open symbols correspond to the experimental data whereas
the lines are fits of the spectra with a strong-coupling model (see text). Black
dots indicate the energy location of the dip in the experimental spectra and the
dotted gray line is a guide to the eye. The energies are measured with respect
to the energy of the negative-bias coherence peak, $\Dp^-$. The spectra in (a)
and (c) are vertically offset for clarity.
}
\end{figure}

For every pixel in a spectroscopic map we assign a gap $\Dpr=(\Dp^+-\Dp^-)/2$,
with $\Dp^+$ and $\Dp^-$ the energy of the positive and negative bias coherence
peaks in the $dI/dV$ spectrum. We find that $\Dpr$ is locally correlated with
the superstructure: in Fig.\ref{fig:fig2}b and \ref{fig:fig2}c the bright
diagonal stripes corresponding to larger $\Dpr$ are located at the maxima of the
surface corrugation. This phenomenon is observed in the three samples studied.
Figure~\ref{fig:fig3}b shows that the gap distribution at constant $\varphi$
follows a sinusoidal dependence on $\varphi$. Averaging all pixels at a constant
$\varphi$ results in a smooth curve with a peak-to-peak amplitude of $8 \pm 2$\%
of the mean value. This amplitude is similar to the $\approx 9$~\% value
reported for Bi-2212 \cite{Slezak-2008}.

A first clue on the variation of the dip energy, $\Ed$, with the gap can be
obtained from averaging spectra with the same $\Dp$. The average spectra shown in
Fig.~\ref{fig:fig2}e indicate that the coherence peaks' height and $\Ed$ follow
a systematic trend with the gap. Figure~\ref{fig:fig3}c shows the same spectra
but with the energy shifted by $\Dp^-$. This representation clearly shows that
$\Wd=|\Ed-\Dp^{-}|$ decreases when the gap increases, in agreement with
tunneling and ARPES data obtained in Bi-2212 by several groups
\cite{Zasadzinski-2001, Damascelli-2003, Hoogenboom-2003b}. Recent STM results
have suggested that $\Wd$ is constant in Bi-2212 \cite{Pasupathy-2008}. However,
the spectra reported in Ref.~\onlinecite{Pasupathy-2008} have a strongly
asymmetric background and a normalization procedure was used to remove this
background before determining $\Wd$. In contrast, no background subtraction is
necessary in our analysis, and $\Wd$ is calculated directly from each of the raw
$dI/dV$ curves. Furthermore, according to the data of Fig.~3B in
Ref.~\onlinecite{Pasupathy-2008}, the claim of a constant $\Wd$ can only be made
with an accuracy of roughly 10\%, which is on the order of the systematic
variation in $\Wd$ that can be clearly seen in our measurements.

Determining the relationship between the gap and the CME requires a reliable
estimation of both quantities from the tunneling spectra. We have fit the
average spectra  of Fig.~\ref{fig:fig2}e using a strong coupling model
\cite{Eschrig-2000} that considers a tight-binding band structure and the
interaction of $d$-wave Bogoliubov quasiparticles with a $(\pi,\pi)$ resonant
mode \cite{Hoogenboom-2003b, Levy-2008}. Figure~\ref{fig:fig3}c shows the
excellent agreement between the fits and the experimental data.  Contrary to
theoretical expectations \cite{Levy-2008}, for all spectra in
Fig.~\ref{fig:fig3}c the CME obtained from the fit falls within 1~meV of
$\Wd=|\Ed-\Dp^{-}|$ calculated from the experimental data. This allows us to
deduce the CME directly from each of the $\sim65000$ spectra in the
spectroscopic maps and thus build $\Wdr$ maps directly from raw data.

\begin{figure}[tb]
\includegraphics[width=\columnwidth]{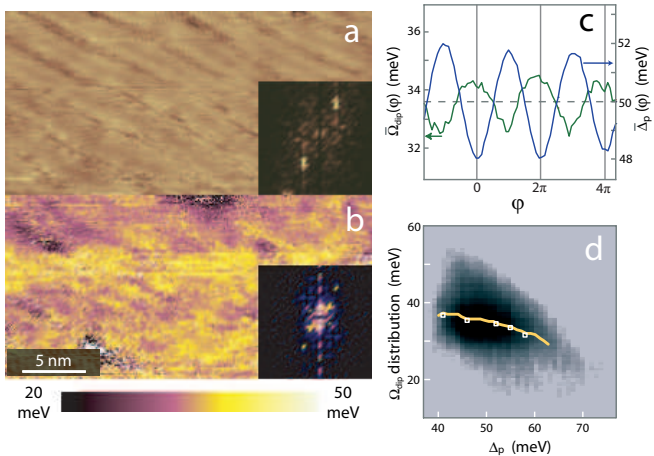}
\caption{\label{fig:fig4}
Spatial modulation of the collective-mode energy (CME) in nearly optimally-doped
Bi-2223. (a) Topographic image of a $30 \times 15$\,nm$^2$ region including the
field of view of Fig.~\ref{fig:fig2}a. Insert: square modulus of the Fourier
transform of the topography depicting the two peaks at the wave-vectors of the
superstructure $\pm 2\pi/a_{s}$. Only the central part of the reciprocal space
is displayed, excluding the four atomic-lattice peaks. (b) Simultaneously
acquired CME map. White pixels indicate spectra where the energy of the dip was
not resolved. Insert: Fourier transform of the CME map displaying the same
region in reciprocal space than in (a). (c) Phase dependence of the CME $\Wdb$
(green line) and gap $\Dpb$ (blue line) averaged at constant $\varphi$. The gap
and the CME are locally \textit{anti}-correlated. (d) Color-scale representation
of $\Wd$ distributions for spectra with the same $\Dp$. The yellow curve
corresponds to the average $\Wd$ as a function of $\Dp$, computed from the
geometrical mean of each histogram. The open squares indicate the CME obtained
from fitting the average spectra shown in Fig.~\ref{fig:fig3}c.
}
\end{figure}

The $\Wdr$ maps exhibit stripes running perpendicular to $\vec{a}_{s}$ (see
Fig.~\ref{fig:fig4}b). The spatial modulation of $\Wdr$ has the same period and
direction as the superstructure, which is confirmed by the two peaks located at
$\pm 2\pi/a_{s}$ in the Fourier transform of the map. We note that
$\Omega(\vec{r})$ maps were reported in Ref.~\onlinecite{Lee-2006} for Bi-2212.
In these maps $\Omega$ was determined from a peak in $d^2I/dV^2$ and no
correlation with the superstructure was observed. Our fits presented here show
that the dip minimum, not the $d^2I/dV^2$ peak, is the appropriate estimation of
$\Omega$. The stripes corresponding to low $\Wd$ in Fig.~\ref{fig:fig4}b
systematically coincide with large gap values. In order to correlate the
$\varphi$-evolution of the CME and the gap we define a $\varphi$-averaged CME,
$\Wdb$, and compare it with $\Dpb$ in Fig.~\ref{fig:fig4}c. The comparison
demonstrates that $\Wdb$ is \emph{anti}-correlated with $\Dpb$. The peak-to-peak
amplitude of the CME modulation is 6\% of the mean value 33.5~meV. This
anti-correlation is also evident from the distribution of $\Wd$ for spectra with
the same $\Dp$ (Fig.~\ref{fig:fig4}d). The average $\Wd$ deduced from the
distribution is in excellent agreement with the CME obtained by fitting the
average spectra with the same $\Dp$.

The evidence presented in this work indicates that the collective mode at the
origin of the dip is the $(\pi,\pi)$ spin resonance detected by inelastic
neutron scattering \cite{Sidis-2004}. First, excellent fits of the STM data over
a wide energy range are obtained only if assuming that the collective mode is
located at $(\pi,\pi)$ \cite{Eschrig-2000, Hoogenboom-2003b, Levy-2008}. Second,
we obtain a CME in the 30--40~meV range, in agreement with the few neutron data
in Bi-based cuprates \cite{Fong-1999}. Third, the CME estimated from STM spectra
decreases on increasing $\Dp$ (Fig.~\ref{fig:fig4}d), and consequently reducing
doping \cite{Kugler-2006}. In the case of underdoped YBa$_2$Cu$_3$O$_{6+x}$, the
only compound for which the doping-dependence of the spin resonance has been
thoroughly characterized, a similar trend is observed \cite{Sidis-2004}. This
must be contrasted with the complete absence of doping dependence found for
optical phonons in Bi-2223 \cite{Boris-2002}. Our fourth piece of evidence is
that the CME is modulated over a lengthscale of $\sim 5a$, implying that the
collective mode must have a similarly short coherence length. Remarkably, the
coherence length of the $(\pi,\pi)$ resonance is only a few unit cells
\cite{Sidis-2004}. This new result, that the CME and the superconducting gap
vary coherently over the short lengthscale of the superstructure, is the key
point of our Letter.

The results presented here have important implications on uncovering the
relevance of this collective excitation to the superconducting state. It has
been reported that the dip feature is neither detected above $T_{c}$ nor within
the vortex cores \cite{Fischer-2007}, and that the intensity of the spin
resonance is negligible at $T> T_{c}$ \cite{He-2001}. These observations support
that the $(\pi,\pi)$ spin resonance has an intimate connection with the
superconducting state. Several theories were proposed in order to explain
high-$T_{c}$ superconductivity, but none has yet managed to generate consensus
in the community. Our results are in agreement with the spin-fluctuation
scenario \cite{Scalapino-1986} that predicts the existence of a $(\pi,\pi)$
resonance \cite{Monthoux-1994} as a consequence of the feedback of pairing on
the spin fluctuations \cite{Chubukov-2003}. Nevertheless, the findings presented
in this letter challenge the theories that do not account for the short
length-scale modulation and the local anti-correlation of the gap and the CME.
Predicting this anti-correlation at the local scale is a critical test for any
scenario aiming to explain high-$T_c$ superconductivity.

We thank A. A. Manuel and A. Petrovi{\'c} for enlightening discussions. This
work was supported by the MaNEP National Center of Competence in Research of the
Swiss National Science Foundation.

\end{document}